
\input harvmac.tex
\overfullrule=0pt

\def\TeV{\rm TeV}
\def\gu{\gamma _{\mu}}
\def\GeV{\rm GeV}

\Title{\vbox{\baselineskip12pt\hbox{BUHEP-94-14}}}
{\vbox{\centerline{Goldstone Bosons in the}
\vskip2pt
\centerline{Appelquist-Terning ETC Model}}}

\centerline{Bhashyam Balaji}
\bigskip\centerline{Physics Department}
\centerline{Boston University}\centerline{Boston, MA 02215}
\vskip .3in

\centerline{\bf Abstract}
\bigskip
It is demonstrated that the extended technicolor model proposed
recently by Appelquist and Terning has  pair of potentially light
 $U(1)$ Goldstone bosons coupling to ordinary matter with strength
$2m_f\over F_{\pi}$, where $m_f$ is the mass of the fermion and
$F_{\pi} \approx 125\,\GeV$. These Goldstone bosons could get a
mass if the spontaneously broken $U(1)$ symmetries are also
explicitly broken, by physics beyond that specified in the model.
 An attempt to
break these symmetries by embedding the model into a larger
gauge group seems to be inadequate. The problem is because
there are too many representations and there
is a mismatch between the number of condensates
and the number of gauge symmetries broken.

\Date{7/94}

\vfil\eject

\newsec{INTRODUCTION}

The reasons for considering the standard model of electroweak
interactions to be incomplete are well known. In particular, the
scalar Higgs sector possesses  unsatisfactory features
such as the naturalness/gauge hierarchy  problem
\ref\KW{K. G. Wilson, {\it Phys. Rev.} {\bf D3}, 181 (1971)},
arbitrariness of Yukawa couplings and triviality
\ref\TRI{R. Dashen and H. Neuberger, {\it Phys. Rev. Lett.} {\bf 50} (1983)
1897}.
These are particularly disturbing since electroweak symmetry
breaking is responsible for endowing ordinary fermions and
the weak gauge bosons with their masses.

One  proposal  for eliminating these problems assumes that there
are no fundamental Higgs scalars.
Instead, one postulates the existence of a new set of
fermions---technifermions---which
interact via a new parity-conserving strong force
called technicolor
\ref\SWLS{S. Weinberg, {\it Phys. Rev.} {\bf D19} (1979) 1277;
L. Susskind, {\it Phys. Rev.} {\bf D20} (1979) 2619 }.
Crudely speaking, `standard'
technicolor  mimics QCD at a
higher energy scale. Walking technicolor
\ref\WTC{B. Holdom, {\it Phys. Rev.} {\bf D24} (1981) 1441; {\it Phys. Lett.}
{\bf 150B} (1985) 301; T. Appelquist, D. Karabali and L. C. R.
Wijewardhana, {\it Phys. Rev. Lett.} {\bf 57} (1986) 957; T. Appelquist and
L. C. R. Wijewardhana, {\it Phys. Rev.} {\bf D36} (1987) 578; K. Yamawaki, M.
Bando
and K. Matumoto, {\it Phys. Rev. Lett.} {\bf 56}, (1986) 1335;
T. Akiba and  T. Yanagida,{\it Phys. Lett.} {\bf 169B} (1986) 432}
modifies relations  obtained by naive
scaling from QCD, due to large anomalous dimensions  of relevant
composite operators. Technicolor enables us to solve part of the problem
---gauge boson mass generation---based on nontrivial dynamics.

The dynamical generation of   fermion masses requires yet another
interaction
---
Extended Technicolor (ETC)
\ref\DS{S. Dimopoulos and L. Susskind, {\it Nucl. Phys.} {\bf B155}
237 }
\ref\EL{E. Eichten, K. Lane, {\it Phys. Lett.} {\bf B90} 125}.
It generates
the  current algebra masses of ordinary fermions  by communicating the
dynamical technifermion masses to the ordinary fermions. In other words,
 ETC gauge bosons couple ordinary fermions to technifermions, and couple
to the various flavors differently.

Three of the Goldstone bosons produced
by the spontaneous chiral symmetry breaking of techniflavor symmetry
 contribute to $W^\pm$ and $Z^0$ masses. `Realistic' models
 usually
contain  other (pseudo-) Goldstone bosons, most of which acquire some mass from
color or electroweak interactions \EL.

An ETC model  explaining the wide range of values of ordinary
fermion masses and the CKM matrix  elements  has been elusive.
In particular, it is  hard to construct models compatible with
experiments. Strong  constraints from
flavor-changing neutral current experiments ruled out QCD-like models
long ago. Moreover, there are unwanted massless goldstone bosons
in any  extended technicolor model with too many fermion
representations \EL .

Recently, severe constraints have also come from precision
electroweak measurements. Assuming the scale of new physics to be large
compared to the $W$ mass, it is found that there are important
corrections to electroweak observables that are `oblique'---i.e.,
correction to gauge boson propagators
\ref\KLY{D. C. Kennedy and B. W. Lynn, {\it Nucl. Phys.} {\bf
B322}, 1 (1989)}.
These oblique corrections
 are encoded in three parameters---S, T and U
\ref\PT{M. Golden and L. Randall, {\it Nucl. Phys.} {\bf B361},3 (1991);
B. Holdom and J. Terning, {\it Phys. Lett.} {\bf B247}, 88 (1990);
M. Peskin and T. Takeuchi, {\it Phys. Rev. Lett. } {\bf 65}, 964 (1990);
A. Dobado, D. Espriu, and M. Herrero, {\it Phys. Lett } {\bf B253}, 161
(1991);
M. Peskin and T. Takeuchi, { \it Phys. Rev.} {\bf D46} (1992) 381;
R. S. Chivukula, M. Dugan and M. Golden, {\it Phys. Lett.} {\bf B292}
(1992) 435}.
The parameter T
\ref\VEL{M. Einhorn, D. Jones, and M. Veltman,
{\it Nucl. Phys.} {\bf B191} (1981) 146;
A. Cohen, H. Georgi, and B. Grinstein {\it Nucl. Phys.} {\bf B232}
(1984) 61}
is a measure of weak-isospin breaking---smallness of T means that
technicolor models with large weak isospin breaking ( needed to generate
the  b-t mass
difference, for instance) are severely constrained. Calculations of the
S parameter in  QCD-like technicolor models with isospin symmetry
indicates that  technicolor models with too many representations
(the one-family technicolor model, for instance) are incompatible
with the experimental value of S.
Of course, those models are already ruled out on other considerations \EL.
However, it is unclear if walking technicolor  models are
incompatible with the experimental values of S and T
\ref\LSU{ R.Sundrum and Stephen D. H. Hsu, {\it Nucl. Phys.} {\bf B391}
(1993) 127;
Markus A. Luty and R. Sundrum, {\it Phys. Rev. Lett.} {\bf 70}
(1993) 529; T. Appelquist and
J. Terning, {\it Phys. Lett.} {\bf B315} (1993) 139}
or corrections to the $Zb\bar b$ vertex
\ref\zbb{R. S. Chivukula,  S. B. Selipsky, and E. H. Simmons,
{\it Phys. Rev. Lett.} {\bf 69} (1992) 575;
R. S. Chivukula,  E. Gates, E. H. Simmons, and
J. Terning, {\it Phys. Lett.} {\bf B311} (1993) 157}
, due to the difficulties inherent in performing reliable
calculations  in a strongly-interacting theory.

An attempt in this direction was made recently by Appelquist
and Terning
\ref\APTE{T. Appelquist and J. Terning, {\bf YCTP-P21-93} {\bf BUHEP-93-23},
hep-ph/9311320}.
They constructed an ETC model and used it to
produce a wide range of fermion masses. They also argued how the model could
be compatible with experimental constraints such as the value of S,
FCNC and small neutrino masses.

In this paper, we will begin by demonstrating the existence  of two
potentially light U(1) Goldstone bosons in the  Appelquist-Terning
model. Appelquist and Terning  only specify gauge interactions below
$1000\,\TeV$
and this is what we  take to be the model.  They also discuss the need
for non-renormalizable operators arising from physics beyond
$1000\,\TeV$. Those operators could  give mass to the two goldstone bosons
by explicitly breaking their chiral symmetries.
We shall discuss such a possibility  assuming  that these
non-renormalizable operators  arise from a gauge theory.
It will be shown that simple extensions, such as
 embedding into larger groups, will not solve the problem.
Our analysis will naturally lead us to the main reason for the
problem---a mismatch between the number of broken diagonal gauge
generators and the number of condensates.
Note that the U(1) Goldstone bosons discussed here are different from
the `$P^0$'  discussed by Eichten and Lane \EL
---those are avoided by
implementing Pati-Salam unification and avoiding repeated representations.

For brevity, the potentially light pseudo-Goldstone bosons will be called
axions.
However, they have nothing to do with the conventional Peccei-Quinn
axion  since they couple to anomaly free currents.
In Section 2, we briefly describe the  model, paying attention to the
symmetry breaking pattern. In Section 3, we  show the existence of
the unwanted massless Goldstone bosons
and present an analysis of the general reasons for the problem.
We discuss the inadequacy of some `natural' ways of addressing the problem
and present a  solution.
Finally, we present our conclusion.

\bigskip
\bigskip

\newsec{\bf The Appelquist---Terning Model}

\bigskip

 The gauge group is taken to be  $SU(5)_{ETC}\otimes  SU(2)_{HC}
\otimes SU(4)_{PS}\otimes SU(2)_L \otimes U(1)_R $ with fermion
content (all fermions taken to be left-handed):

$$\eqalign{\psi _1 & = (5,1,4,2)_0 \cr
\psi _2 & = (\overline 5 ,1,\overline 4 ,1 )_1 \cr
\psi _3 & = (\overline 5 , 1,\overline 4 ,1 )_{-1} \cr
\psi _4 & = (1,1,6,1)_0 \cr} \qquad
\eqalign{\psi _5 & = (1,2,6,1)_0 \cr
\psi _6 & = (10,1,1,1)_0 \cr
\psi _7 & = (5,1,1,1)_0 \cr
\psi _8 & = ( \overline {10} ,2,1,1)_0 \cr}\eqno(1) $$

\noindent
Here $SU(2)_{HC}$ is an additional strong gauge group which is
needed to help break the $SU(5)_{ETC}$ down to
$SU(2)_{TC}$. Hypercharge, $Y$,(normalized by $Q=T_{3L} + Y/2$) is
given by $Y=Q_R + T_{15} ^{PS}$, where $T_{15} ^{PS} = $
diag($1/3,1/3,1/3,-1$)
is a generator of the $SU(4)_{PS}$
Pati-Salam group which implements quark-lepton unification
(For details regarding the motivation for the choice of the gauge group
and fermion representation content see \APTE).

At the Pati-Salam breaking scale (taken to be around $1000\,\TeV$ ), a
condensate is assumed to form in the channel  $ (\overline 5 ,1,
\overline 4 ,1)_{-1} \times (5,1,1,1)_0 \rightarrow (1,1,\overline 4
, 1)_{-1} $. This channel ($\langle\psi _3\psi _7\rangle
\neq0$) is not the most attractive Channel (MAC)
\ref\MAC{S. Dimopoulos, S. Raby, and L. Susskind, {\it Nucl. Phys.} {\bf
B169} (1980)373}.
Instead, new physics is presumed to trigger its formation. This  condensate
breaks the $U(1)_R$ and  $SU(4)_{PS}$ gauge groups leading to the gauge group
$SU(5)_{ETC}\otimes SU(2)_{HC}\otimes SU(3)_C \otimes SU(2)_L
\otimes U(1)_Y$ below $\Lambda _{PS}$.

Next, it is assumed that at $\Lambda _5 \approx 1000\,\TeV$ a condensate
forms in the channel $ (10,1,1,1)_0\times (10,1,1,1)_0 \rightarrow
(\overline 5 , 1,1,1)_0$ ---i.e., $\langle\psi _6\psi _6\rangle\neq0$ .
 The singlet channel $10\times {\overline {10}} \rightarrow 1$ is
disfavored as $SU(2)_{HC}$ is assumed to be relatively strong, so as
to resist breaking. Then, the condensate
$(\overline 5 ,1,1,1)_0$ breaks the gauge symmetry to $
SU(4)_{ETC}\otimes SU(2)_{HC} \otimes SU(3)_C \otimes SU(2)_L \otimes
U(1)_Y$.

The next condensation is the attractive channel
$(\overline 4 ,2,1,1)_0 \times (6,2,1,1)_0 \rightarrow (4,1,1,1)$
--- $\langle\psi _8\psi _8\rangle _1\neq 0 $ ---
 which
is taken to occur at  $\Lambda _4\approx 100\,\TeV$.
( The condensate is subscripted to distinguish this channel
from  another at $\Lambda _3$  arising from a different
piece in $\psi _8$.)
The so-called `big MAC' criterion
is applied here. When two or more relatively  strong gauge
interactions are at play, the favored breaking channel is
determined by the sum of the interactions. It is a
generalization of the ordinary MAC criterion. Hence, below $\Lambda _4$
the gauge group is $SU(3)_{ETC}\otimes SU(2)_{HC}\otimes
SU(3)_C \otimes SU(2)_L \otimes U(1)_Y$.

The final stage of ETC breaking takes place at the scale $\Lambda _3
\approx 10\,\TeV$ with the big MAC condensate $(\overline 3 ,2,1,1)_0
\times (\overline 3 ,2,1,1)_0\rightarrow (3,1,1,1)_0$\
--- $\langle\psi _8\psi _8\rangle_2\neq 0$.
This breaks $SU(3)_{ETC}$ to $SU(2)_{TC}$ so that the gauge group below
$\Lambda _3 $ is $SU(2)_{TC}\otimes SU(2)_{HC}\otimes SU(3)_C \otimes
SU(2)_L\otimes U(1)_Y$. Hypercolored  paricles are confined at
$\Lambda _{HC} \approx \Lambda _3$, and the HC sector decouples from
ordinary fermions and technifermions. We then have a  one-family
technicolor model ( actually there is an additional ``vector'' quark
\APTE).

Finally, at the technicolor scale $\Lambda _{TC}$, the $SU(2)_{TC}$
becomes strong resulting in  condensation in the $2\times 2
\rightarrow 1 $ channel--- $\langle\psi _1\psi _2\
\rangle\neq 0$,  $\langle\psi _1\psi _3\rangle\neq 0 $
and  $\langle\psi _1\psi _6\rangle\neq 0$.
This breaks the  electroweak gauge group  $SU(2)_L\otimes U(1)_Y $  to
 $U(1)_{\rm em}$.

\newsec{\bf The Axion Analysis }

\bigskip

For each fermion representation, there is a global $U(1)$ symmetry
current. In general, these currents have gauge anomalies. From these
currents, one can form anomaly-free combinations; these correspond to
exact global symmetries of the theory.  One then follows
this global symmetry through the various gauge symmetry breakings and
investigates
whether or not it is spontaneously broken at any scale.
At each stage of symmetry breaking, one forms linear combinations
of the gauged and global currents  that leave the condensates invariant.
These remaining symmetries generate
unbroken global $U(1)$'s. The remaining orthogonal combinations
couple to the massive gauge bosons
\ref\KL{K. Lane, {\it Proceedings of the The Fifth Johns Hopkins
Workshop on Current Problems in Particle Theory, Baltimore, Maryland} 1981}.

To begin with, there are   eight gauge singlet global $U(1)$
currents $j^A _{\mu} =  \overline {\psi _A}  {\gamma _\mu} {\psi _A}$ ,
where  $A=1,...8$ corresponding to the eight representations. Each of
them has an anomalous divergence due to  the strong ETC, PS and/or HC
interactions.

$$\eqalign{D_1& = 8S_5+10S_4\cr
D_2& = 4S_5+5S_4\cr
D_3& = 4S_5+5S_4\cr
D_4& = 2S_4\cr
D_5& = 6S_2+4S_4\cr
D_6& = 3S_5\cr
D_7& = S_5\cr
D_8& = 6S_5+10S_2\cr}\eqno(2)$$

\noindent
Here  $ D_A = {\partial }_{\mu} j^{\mu} _A$ and $S_n = {g_n ^2 \over
32{\pi }^2}F_n \cdot  \tilde{F_n}$, where  $n=2,4,5$ correspond to
to gauge groups $SU(2)_{HC}$, $SU(4)_{PS}$ and $SU(5)_{ETC}$ respectively.
The electroweak $SU(2)_L$ instanton has a negligible effect  around  $\Lambda
_{PS}$, and weak anomalies are ignored.

{}From these currents, one can form five
gauge anomaly free symmetry currents, which are

$$\eqalign{J^1 _{\mu} & =  j^1 _{\mu} - j^2 _{\mu} - j^3 _{\mu} \cr
 J^2 _{\mu} & = j^6 _{\mu} - 3j^7 _{\mu} \cr
 J^3 _{\mu} & =  j^1 _{\mu} - 5j^4 _{\mu} - 8j^7 _{\mu} \cr
 J^4 _{\mu} & =  -10j^4 _{\mu} + 5j^5 _{\mu} +6j^6 _{\mu} - 3j^8 _{\mu} \cr
 J^5 _{\mu} & =  -j^2 _{\mu} +  j^3 _{\mu}\cr}\eqno(3)$$

\noindent
One of these currents, $J^5 _{\mu}$, is actually the $U(1)_R $ gauge current.
So
the global symmetry above $\Lambda _{PS}$ is $U(1)^4$.
(To include the $SU(2)_L$ we take three of the four linearly independent
current combinations not containing $j^1 _{\mu}$. The symmetry would be
$U(1)^3$; this will not change our conclusions. Note that the $U(1)_R$
anomaly is irrelevant for our considerations  due to the absence of
instantons.) We follow this
global symmetry through all of the gauge symmetry breakings; spontaneous
breakdown of any global symmetry would result in a corresponding Goldstone
boson at that scale.

 Below $\Lambda _{PS}$ one can form four
combinations from the above five currents which leave invariant the gauge
symmetry breaking condensate $\langle\psi _3 \psi _7\rangle
\neq0$---the fifth one being the $U(1)_R$ gauge current [5]. So there still
exists a $U(1)^4$ symmetry generated by ( keeping the same symbol,$J$,
for these currents )

$$\eqalign { J^1 _{\mu} & = -3j^1 _{\mu} + 3 j^2 _{\mu} + 3j^3 _{\mu} + j^6
_{\mu} -
 3j^7 _{\mu}\cr
 J^2 _{\mu} & = -7j^1 _{\mu} + 8j^2 _{\mu} + 8j^3 _{\mu} -5j^4 {\mu}
 - 8j^7 _{\mu} \cr
 J^3 _{\mu} & = j^1 _{\mu} - 2j^2 _{\mu}  \cr
 J^4 _{\mu} & = -10j^4 _{\mu} + 5j^5 _{\mu} +6j^6 _{\mu} - 3j^8
_{\mu}\cr}\eqno(4)$$

One can  proceed to find the four global symmetry currents,
which  remain conserved below $\Lambda _5$, to be

$$\eqalign{J^1 _{\mu} & = 18 ( j^1 _{\mu} - j^2 _{\mu} - j^3 _{\mu}) -10j^4
_{\mu} +
 5j^5 _{\mu} + 18j^7 _{\mu} - 3j^8 _{\mu}\cr
 J^2 _{\mu} & = -10j^4 _{\mu} + 5j^5 _{\mu} + 6j^6 _{\mu} -3j^8 _{\mu}
 - 6J^{5E} _{\mu} \cr
 J^3 _{\mu} & = j^1 _{\mu} - 2j^2 _{\mu}  \cr
 J^4 _{\mu} & = -7j^1 _{\mu} + 8j^2 _{\mu} + 8j^3 _{\mu} - 5j^4 _{\mu}
 -8j^7 _{\mu}\cr}\eqno(5)$$

\noindent
Here  $J^{5E} _{\mu}$ is the gauge current corresponding to the diagonal
generator in SU(5) which is not in SU(4). Note that $J^2 _{\mu}$ is left
unbroken although the gauge and global currents are separately broken.
This is analogous to the situation in the standard model where
$SU(2)_L \otimes SU(2)_R \rightarrow SU(2)_V$.
The $SU(5)_E$ generator in the fundamental representation
 is chosen to be ${\rm diag } {1\over 2}
(-4,1,1,1)$ ---this automatically fixes the U(1) charges in the other non-
fundamental representations. This is merely a convenient choice of a
basis.

One can likewise obtain the global symmetry currents below $\Lambda _4$
and $\Lambda _3$.
Finally, one investigates the  $U(1)$  global symmetries below the
electroweak symmetry breaking scale. This entails  three condensates ---
  $\langle\psi _1\psi _2\rangle\neq0 $,\
  $\langle\psi _1\psi _3\rangle\neq0 $  and
  $\langle\psi _1\psi _6\rangle\neq0 $.
Only two of the four global symmetries are  realized in the Wigner-Weyl
mode,
namely those corresponding to

$$\eqalign{J^1 _{\mu} & = - 36 ( j^1 _{\mu} - j^2 _{\mu} - j^3 _{\mu}) +10j^4
_{\mu}
  - 25 j^5 _{\mu} - 18j^6 _{\mu} \cr
     &\qquad- 36j^7 _{\mu} + 15 j^8 _{\mu}
  + 8J^{5E} _{\mu} - 24J^{4E} _{\mu} + 6J^{3E} _{\mu} \cr
\noalign{\hbox{ }}
 J^2 _{\mu} & = -18j^1 _{\mu} + 18j^2 _{\mu} + 20j^3 _{\mu} - 15j^4 _{\mu}
  - (5/3)j^5 _{\mu} + 2j^6 _{\mu} \cr
    &\qquad-20 j^7 _{\mu} + j^8 _{\mu}
  - 2J^{5E} _{\mu} + 2J^{3E} _{\mu} + J^{SU(2)_3} _{\mu}\cr}\eqno(8)$$

\noindent
The $SU(2)_3$ generator in the fundamental representation
is normalized as ${\rm diag }(1,-1)$. The $SU(4)$ and $SU(3)$ generators
in the fundamental representation are normalized as ${\rm
diag}(-3,1,1,1)$ and ${\rm diag}(-2,1,1)$ respectively.
Note that the $j^{\rm em} _{\mu }$ piece is
irrelevant for our purposes since $U(1)_{\rm em}$ is unbroken.

The other two broken global $U(1)$ currents ( up to unbroken pieces )
are

$$\eqalign{\
J^3 _{\mu} & = -36 ( j^1 _{\mu} - j^2 _{\mu} - j^3 _{\mu}) - 30j^4
_{\mu}
  - 5j^5 _{\mu}+ 6j^6 _{\mu}\cr
  &\qquad- 36j^7 _{\mu} + 3j^8 _{\mu} - 6J^{5E} _{\mu} + 6J^{3E}
_{\mu}\cr
\noalign{\hbox{ }}
J^4 _{\mu} & = -70j^1 _{\mu} + 74 j^2 _{\mu} + 72 j^3 _{\mu}  - 55j^4 _{\mu}
  + (20/3) j^5 _{\mu} + 8j^6 _{\mu}\cr
   &\qquad- 72 j^7 _{\mu} + 4j^8 _{\mu}
  - 8J^{5E} _{\mu} + 8J^{3E} _{\mu}\cr}\eqno(9)$$

One can incorporate the effect of the $SU(2)_L$ anomaly  by
constructing three linearly independent currents(from equations $(8)$
and $(9)$) not
containing $j^1 _{\mu}$. The unbroken current is

$$\eqalign{
{\bf J}^0 _{\mu} & = J^1 _{\mu} - 2 J^2 _{\mu}}\eqno(10)$$
\noindent
The remaining two (linearly independent) spontaneously broken
global $U(1)$ currents are

$$\eqalign{\
{\bf J}^1 _{\mu} & = J^3 _{\mu} - J^1 _{\mu}  \cr
{\bf J}^2 _{\mu} & = {35\over 18} J^1 _{\mu} -  J^4 _{\mu}\cr}\eqno(11)$$

In terms of the first generation ordinary fermions, the two spontaneously
broken global
currents are

$$\eqalign{
 {\bf J}^1 _{\mu} & = -28( \overline {q^1 _L}\gu q^1 _L+
\overline {l^1 _L} \gu {l^1 _L}  - \overline u^c _R \gu u^c _R -
\overline {d^c _R} \gu d^c _R - \overline {e^c _R} \gu e^c _R )
+ \cdots \cr
\noalign{\hbox{ }}
{\bf J}^2 _{\mu} & = -4\overline {u^c _R}\gu u^c _R  -2(\overline
{d^c _R}\gu d^c _R  + \overline {e^c _R} \gu e^c _R )\cr
      &\qquad-{424\over 9} ( \overline {q^1 _L}\gu q^1 _L+
\overline {l^1 _L} \gu {l^1 _L}  - \overline u^c _R \gu u^c _R -
\overline {d^c _R} \gu d^c _R - \overline {e^c _R} \gu e^c _R) + \cdots
\cr}\eqno(12)$$

\noindent
There is no flavor mixing in this model, the CKM angles are assumed to arise
from
higher  dimensional operators.

Our analysis has demonstrated that there are two potentially massless Goldstone
bosons in the model. They couple
with strength $2m_f/{F_{\pi}}$ where $F_{\pi} = 125{\rm GeV}$  to
light fermions of current algebraic mass $m_f$ and are experimentally
ruled out
\ref\AXI{T. W. Donnelly {\it et. al.}, {\it Phys. Rev.} {\bf D18}, (1978)
1607; S. Barshay {\it et. al.}, {\it Phys. Rev. Lett.} {\bf 46}, (1981);
A. Barroso {\it et. al.}, Phys. Lett. {\bf 106B}, (1981) 91; and
R.D. Peccei, in {\it Proceedings of Neutrino'81}, Honololu, Hawaii,
Vol.1, p.149 (1981)}.

Let us assume for the moment that these chiral symmetries are
explicitly broken by physics beyond $1000\,{\rm TeV}$.
It is possible to obtain a rough estimate of the axion
mass ($m_A$)  from four fermion operators generated by
heavy physics. Dashen's formula coupled with vacuum insertion
approximation  yields,

$$\eqalign{
F^2 _{\pi} m^2 _{A} \approx  {\langle \overline T T\rangle ^2 _{\Lambda _S}
\over F^2_S}}\eqno(13)$$

\noindent
where $g_S F_S = \Lambda _S $  is the scale of new physics.
Here $\langle \overline T T \rangle_{\Lambda _3} = 4\times 10^8\GeV ^3$
 is a rough estimate of the technifermion condensates
 used by Appelquist and Terning.
  Quoting  values of the relevant anomalous dimensions from \APTE , we obtain

$$\eqalign{
\langle \overline T T\rangle _{\Lambda _S} \approx
 {<{\overline T}T >_{\Lambda_3} }
\left( {{100 \,{\rm TeV}}\over{10 \, \rm{TeV}}}\right)^{0.67}
\left( {{1000 \,{\rm TeV}}\over{100 \,{\rm TeV}}}\right)^{0.32}
\left( {{\Lambda_S\TeV}\over{1000\,\TeV}}\right)^{\gamma_S}}\eqno(14)$$

\noindent
where $\gamma_S$ is the anomalous dimension from $1000\,\TeV$  to
$\Lambda_S$. Hence

$$\eqalign{
m_A \approx 32 {\left(  {10^3}\TeV \over F_S \right)}^{({1-\gamma _S})}
\GeV}\eqno(15)$$

\noindent
Here $g_S$ is taken to be $O(1)$ and $\gamma _S $ --- a crude estimate
of walking effects between $\Lambda_S$ and $\Lambda _{PS}$ ---is
also expected to be of $O(1)$.

In a $p\bar p $ collider, these neutral `axions'($A_{1,2}$) may be
produced singly via gluon fusion (predominantly)
or by quark-antiquark fusion
\ref\EHLQ{E. Eichten, I. Hinchliffe, K. Lane and C. Quigg,
{\it Rev. Mod. Phys.} {\bf 56} (1984) 579}.
Note that there is not enough phase space for it to be produced in the
pair technipion production mode ( $W^{\pm} \rightarrow A_i P^{\pm}$ ),
where $P^{\pm}$ is the charged pseudo-goldstone boson in the one-family
technicolor model orthogonal
to the `eaten' goldstone bosons, as the mass of the $P^{\pm}$ is expected
to be in excess of $50\,\GeV$ \WTC .
These axions would decay into fermion-antifermion pair, principally
the heavier ones ($b\bar b$ and $\tau\bar \tau$). However, these decays
are not of any experimental significance due to the smallness of
$m_{b,\tau}\over F_{\pi}$.
Hence, a $32\,\GeV$ neutral, color-singlet
`axion'  is unlikely to be detected  in the near future.

A natural approach for  creating these four-fermion operators  would be
to embed the Appelquist-Terning gauge group
into a larger gauge group.  This will break some of the chiral
symmetries so that the axions are no longer strictly massless. The simplest
possible extension ( minimum fermion content and
maximum chiral symmetry breaking)
is to the gauge group
$SU(9)\otimes SU(2)_{HC}\otimes SU(2)_L \otimes U(1)_R$ ; i.e.,
unification of the Pati-Salam and ETC gauge groups. The $SU(9)$ is
assumed to break into $SU(5)_{ETC} \otimes SU(4)_{PS}$. The
minimal representation content yielding us the eight  representations
in the Appelquist - Terning model , under decomposition,  is

$$\eqalign{\
\Psi _1 & =(36,1,2)_0 = (10,1,1,2)_0\oplus (5,1,4,2)_0\oplus (1,1,6,2)_0 \cr
\noalign{\hbox{ }}
\Psi _2 & =(36,1,1)_0 = (10,1,1,1)_0\oplus (5,1,4,1)_0\oplus (1,1,6,1)_0 \cr
\noalign{\hbox{ }}
\Psi _3 & =(\overline {36} , 2, 1 )_0 = (\overline {10} ,2,1,1)_0\oplus
(\overline 5 ,2,\overline 4 ,1 )_0\oplus ( 1,2,6,1)_0  \cr
\noalign{\hbox{ }}
\Psi _4 & = (\overline {36} ,1 ,1 )_{-1} = (\overline 10 ,1,1,1)_{-1}\oplus
(\overline 5 ,1,\overline 4 ,1)_{-1}\oplus (1,1,6,1)_{-1} \cr
\noalign{\hbox{ }}
\Psi _5 & = (\overline {36} ,1,1)_1 = ( \overline 10 ,1,1,1)_1 \oplus
(\overline 5, 1,\overline 4 ,1)_1\oplus (1,1,6,1)_1 \cr
\noalign{\hbox{ }}
\Psi _6 & = (9,1,1,1)_0  = (5,1,1,1)_0\oplus (4,1,1,1)_0\cr
\noalign{\hbox{ }}
\Psi _7 & = (\overline 9 ,1,1)_0 = (\overline 5 ,1,1,1)_0\oplus (\overline 4
,1,1,1)_0\cr
\noalign{\hbox{ }}
\Psi _8 & = (126 ,1 ,1 )_0 \cr
        &= (\overline 5,1,1,1)_0\oplus (1,1,1,1)_0\oplus
(10,1,6,1)_0\oplus (\overline 10 , 1,4,1)_0                \cr
 &\qquad\oplus (5,1,\overline 4 ,1)_0\quad\cr}\eqno(16)$$

Since we want to avoid  Non-Abelian Goldstone bosons,
 no  representations  participating in any condensate should be repeated.
 The original fermions are contained in $\Psi _1$,
...,$\Psi _6$. The representations $\Psi _7$ and $\Psi _8$ are needed to cancel
gauge anomalies.

However, this is not a resolution. While it explicitly breaks some
chiral symmetries, the additional representations create new chiral
symmetries. The `axions' are still there.

Another approach  for tackling this problem would be to gauge the
 two $U(1)$'s that are spontaneously broken at the
electroweak scale.(However, this would require a new set of spectator
fermions so that the theory is anomaly free. One needs to assume that
they are singlets under all but these two $U(1)$ groups and get a mass
from heavy physics.)
Then,  the  two massless $U(1)$ gauge bosons
combine with the two massless goldstone bosons
 to give the  former their masses. The arbitrariness
of the $U(1)$ gauge coupling  can push the gauge boson masses beyond
their current experimental limits.  $Z^{0'}$ mass (lower)bounds
indicate that the U(1) gauge couplings would have to be rather large to
be compatible with experiment.

Another  possibility would be  to arrange to have more than one
condensate at  higher energy gauge symmetry breaking scales.
As a result, the $U(1)$'s  are spontaneously broken at higher energies
and hence are more weakly coupled to the light fermions (weak enough
to be undetected). This way one ensures that there are no
light goldstone bosons at the electroweak scale. However, `$f_{\pi}$'
would have to be between $10^{10}\,\GeV$ and $10^{12}\,\GeV$ from cosmological
\ref\up{J. Preskill {\it et. al.}, {\it Phys. Lett.} {\bf 120B}, (1983)
127; L. F. Abbott, and P. Sikivie, {\it Phys. Lett.} {\bf 120B}, (1983) 133;
and M. Dine and W. Fischler,{\it Phys. Lett.} {\bf 120B} (1983) 137}
and astrophysical considerations
\ref\low{D. A. Dicus {\it et al.}, {\it Phys. Rev.} {\bf D18} (1978);
M. Fukugita, S. Watamura, and M. Yoshimura, {\it Phys. Rev. Lett.} {\bf 48},
(1982) 1522, also {\it Phys. Rev.} {\bf D26}, (1982) 1840; D. S. P. Dearborn
{\it et al.}, {Phys. Rev. Lett.} {\bf 56} (1986) 26; and G. G. Raffelt,
{\it Phys. Rev.} {\bf D33}, (1986) 897}.

The above discussion can be stated in more general terms.
A reason for the problem is well known---too many
representations \EL.
If the number of irreducible  representations ($n_D$) is less than or
equal to the number of simple non-abelian gauge group factors ($n_S$) there
will be no anomaly-free global U(1) currents to worry about.

If $n_D > n_S$, we have $(n_D - n_S)$ anomaly-free global U(1)
symmetries. We then investigate the fate of these symmetries as we pass
through the various gauge symmetry breaking scales.
Suppose the gauge symmetry breaking at $\Lambda $ involves  $c$
condensates and there are $d$ broken diagonal generators which act as
the number operator in the subspace of the fermions in the condensate.
(For instance,
when $SU(n)$ breaks into $SU(n-1)$, the element of the Cartan subalgebra
in $SU(n)$ but not in $SU(n-1)$ acts as the number operator in the $1$ and
$n-1$ subspace separately.) Consider the general linear combination of
these $(n_D - n_S + d)$ $U(1)$ currents
( $J=\sum A_r J^r_\mu$, $r=1,\cdots (n_D -n_S +d)$ )
and tabulate the charges
$Q_i$, ($i=1,\cdots c$) of the $c$
condensates. (Without loss in generality, we are assuming no global
symmetry breaking above this scale.)  Note that these $Q_i$ are linear
combinations of $(n_D - n_S +d)$ variables $A_r$.

The conditions that the charges of the $c$ condensates
vanish can be stated as a set of $c$ linear homogeneous equations in
$(n_D - n_S +d)$ variables. If the rank of the coeficient matrix
($c'$) is less than $c$, only $c'$ condensates are said to have linearly
independent charges. We focus on the $c'$ linearly
independent condensates i.e., those condensates with linearly
independent charges. Only when $c' = d$ are the initial $U(1)^{(n_D - n _G)}$
global symmetries preserved; otherwise there are $c' - d$
exactly massless Goldstone bosons, which would be unwelcome if they
couple to the ordinary fermions. Gauging these $U(1)$'s is
a solution, but this entails the introduction of spectator fermions.

We know of no simple way of
determining $c'$ other than to explicitly construct the global
symmetry currents at each stage of the gauge symmetry breaking.
In any case, the analysis needs to be carried out only when
$c > d$ ; i.e., when the number of condensates is larger than the
number of `diagonal'
gauge generators spontaneously broken at that scale. In particular,  there
could be
many  ( unrepeated ) ETC representations as long as $c \leq d$.

\bigskip
\bigskip

\newsec{\bf CONCLUSION}
\bigskip

We have shown the existence of two potentially light Goldstone bosons in
the Appelquist-Terning ETC model. The problem is related to the fact that
there are too many representations and the number of condensates exceeds
the number of diagonal gauge symmetries broken. These axions  could get
a mass from physics at higher energy scales. An
attempt to break these symmetries with new gauge interactions
seems to be inadequate.

\newsec{\bf Acknowledgements}
\bigskip

I thank  K.Lane for guiding the investigation and for a critical
reading of the manuscript. I also thank R.S.Chivukula and J.Terning for useful
discussions and suggestions.
This work was supported in part by the Department of Energy under
Contract  No. DE-FG02-91ER40676 and by the Texas National Research
Laboratory Commission under Grant No. RGFY93-278.

\bigskip
\bigskip

\listrefs
\bye